\documentclass[11pt]{evn}
\usepackage[T2A]{fontenc}
\usepackage[cp1251]{inputenc}
\usepackage{epsf}
\usepackage[russian,english]{babel}
\usepackage{graphicx}
\long\def\maintitle#1{{\vskip 20mm \begin{center}\section*{#1}\end{center}\nopagebreak[4]}}

\long\def\author#1{{\begin{center}\normalsize{\bf#1}\end{center}\vskip-1em\index{#1}}\nopagebreak[4]}
\long\def\address#1{{\begin{center}\small\noindent#1\end{center}\vskip-8mm}\nopagebreak[4]}

\begin{document}
\noindent\mbox{\small The 13$^{th}$ EVN Symposium \& Users Meeting Proceedings, 2016}

\maintitle{MULTIPLE SIMULTANEOUS GAMMA-RAY EMISSION SITES IN 3C\,84}

\author{Jeffrey~A.~Hodgson$^{1}$, S.S.~Lee$^{1}$, B.~Rani$^{2}$ and J.C.~Algaba$^{1}$}

\address{$^{1}$Korea Astronomy and Space Science Institute\\$^{2}$NASA Goddard, United States}

\begin{abstract} The mis-aligned active galactic nuclei (AGN), 3C\,84, has been observed approximately monthly at four frequencies simultaneously (22\,GHz -- 129\,GHz) on the Korean VLBI Network (KVN) since 2013 as part of the interferometric monitoring of $\gamma$-ray bright AGN (iMOGABA) program. 3C\,84 is known to have to sites of bright radio emission, with the first thought to be near the central super-massive black hole and the second in a slowly moving emission feature, several milliarcseconds south of the core. Analysis of this data suggests that the highly variable $\gamma$-ray emission originates near the SMBH and the slowly rising $\gamma$-ray emission originates in the slow moving feature.
\end{abstract}

\section{Introduction}

The well-studied nearby (z=0.0177) misaligned AGN 3C\,84 is a unique source because unlike many other similar sources, it is detected at $\gamma$-ray energies and it is thought that we are directly observing the jet launching region \cite{1}. One of the biggest unresolved questions in AGN studies is the site of $\gamma$-ray emission in jets. $\gamma$-rays are thought to be produced by the up-scattering of lower energy photons that are either seeded internally to the jet (e.g. \cite{2}) or externally perhaps in the broad line region (e.g. \cite{3}). Evidence has been conflicting, although with most evidence in blazars suggesting the internal mechanism, occurring downstream of the central SMBH (e.g. \cite{4}). The aim of the study is to use simultaneous multi-frequency VLBI observations to determine the location of $\gamma$-ray flaring. 

\section{Observations}

Observations were conducted approximately monthly from January 16 2013 until March 1 2016 at 22\,GHz (14\,mm), 43\,GHz (7\,mm), 86\,GHz (3\,mm) and 129\,GHz (2\,mm) on the Korean VLBI Network (KVN). Data were observed as part of the iMOGABA program \cite{5} and data were reduced using the KVN Pipeline \cite{6}, which is referred to for more details. In total, 32 epochs were observed, with five epochs lost due to technical or weather issues. Data in the $\gamma$-ray bands were obtained from the Fermi-LAT spacecraft and binned monthly. Details of $\gamma$-ray data reduction are in the paper that is being prepared, but similar procedures can be found in \cite{7}. 

\section{Results}

Recent VLBI maps show that the source has two main bright components, the northernmost labelled C1 and the southernmost labelled C3. Setting C1 as a reference, we find motion detected at all observing frequencies except 2\,mm, consistent with previous published results \cite{8}. 


\begin{figure}[ht]
\label{LCs}
\centering
\includegraphics[width=1.4\textwidth, angle=90]{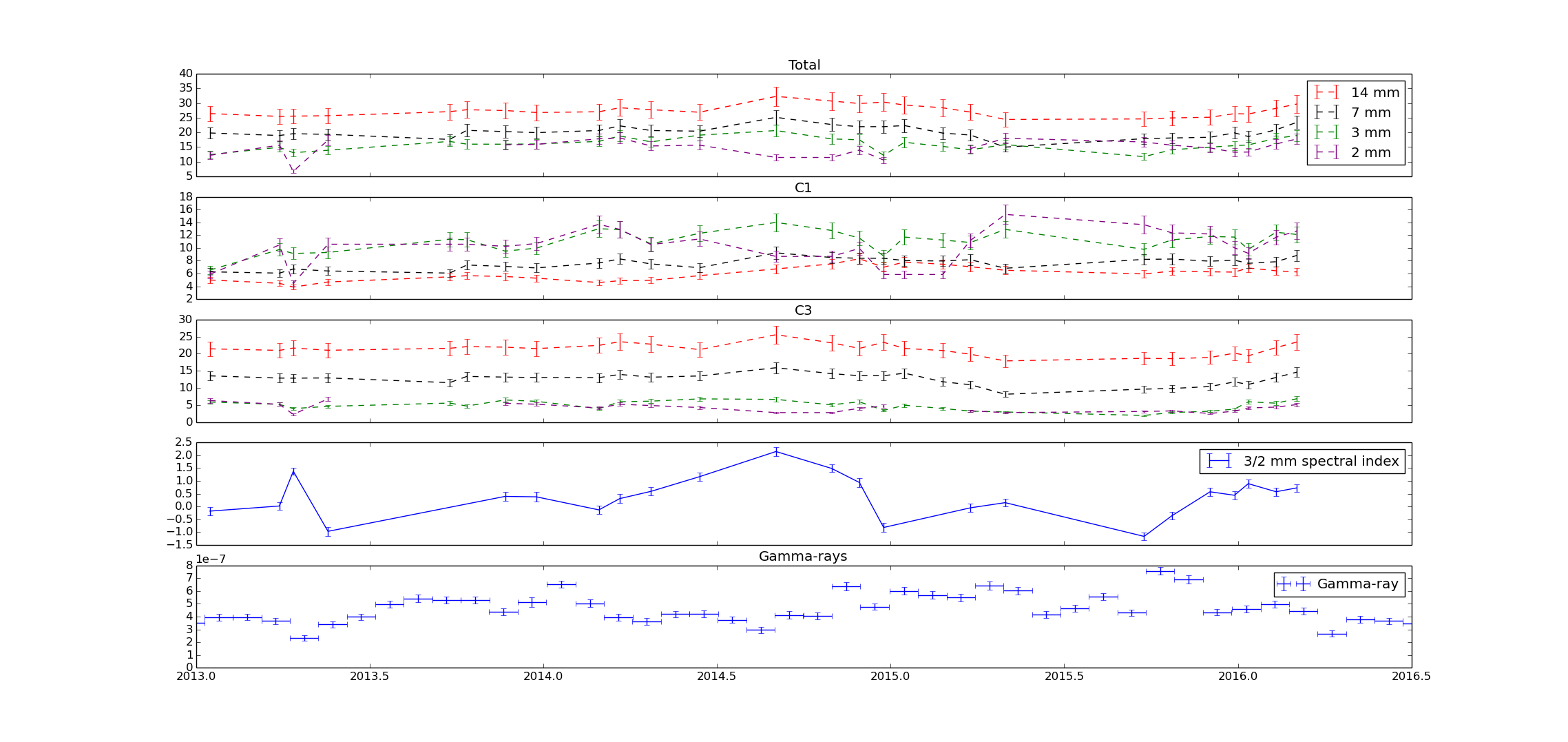}
\caption{Figure 1. X-label: epoch. Y-labels from top to bottom: Total radio flux densities (Jy), C1 flux density (Jy), C3 Flux density (Jy), 3/2 mm spectral index in C1, monthly average $\gamma$-ray flux density (photons/cm$^{2}$/s).}
\end{figure}

In order to determine the relative changes in flux densities consistently across the frequency (and hence resolution) range, the maps were model-fitted using a fixed Gaussian with a full-width-half-maximum of 0.8\,mas and with only two components. While this may affect the absolute accuracy of the flux densities measured, it should improve the relative accuracy. The light-curves from the source are shown in Fig. 1. The top panel shows the light curves at each frequency for the total (C1+C3) flux density. The next panel below shows the C1 decomposed light-curve and the panel below that shows the C3 decomposed light-curve. The next panel shows how the 3\,mm -- 2\,mm spectral index in C1 changes with time and the bottom panel shows the monthly binned $\gamma$-ray flux densities.

We see that the the spectral index appears to be anti-correlated with the $\gamma$-ray light-curve. However, there also appears to be an increasing trend in the $\gamma$-ray flux density that is best reflected with the trend in C3. The spectrum of C1 is highly variable between 3\,mm and 2\,mm, while the spectral shape of C3 is quite consistent across all epochs.

\section{Discussion}

The relatively rapid $\gamma$-ray flaring appears to be correlated with the mm-wave flaring in C1. Since C1 is considered likely to be at or near the location of the central SMBH, this places that site of $\gamma$-ray flaring much closer to the central engine than many other studies have found, where they place the site many parsecs downstream (e.g. \cite{4,7}). Emission found downstream is normally interpreted as due to the seed photons for $\gamma$-ray production originating within the jet. The flaring in C3 could be consistent with this interpretation, but the emission from C1 suggests that the seed photons could potentially originate in the BLR. This therefore suggests that both mechanisms may be responsible for $\gamma$-ray emission, even in the same source. 

\section{Conclusion}

We find evidence for $\gamma$-ray flaring to be occurring in multiple locations within 3C\,84 at the same time. Short time-scale flaring appears to be correlated with mm-wave flaring in the core region, near the SMBH. Longer time-scale $\gamma$-ray emission appears to be associated with a travelling shock several parsecs downstream of the SMBH. We interpret these results as likely being evidence of multiple $\gamma$-ray emission mechanisms being present within 3C\,84.




\begin{thebibliography}{5}
\bibitem{1}
{\it Nagai H. et al.} VLBI Monitoring of 3C 84 (NGC 1275) in Early Phase of the 2005 Outburst // ~-- 2010. ~-- PASJ, 62L, 11N
\bibitem{2}
{\it Band, D. L., Grindlay, J. E.} The synchrotron-self-Compton process in spherical geometries. II - Application to active galactic nuclei // Astrophysical Journal ~-- 1986.~-- 308, 576B
\bibitem{3}
{\it 
Sikora, M., Madejski, G.}  Blazars //  American Institute of Physics Conference Series.~-- 2001.~-- 558, 275
\bibitem{4}
{\it Jorstad S. et al.} Polarimetric Observations of 15 Active Galactic Nuclei at High Frequencies: Jet Kinematics from Bimonthly Monitoring with the Very Long Baseline Array // AJ  -- 2005. -- 130, 1418
\bibitem{5}
{\it Algaba, J.-C et al.} Interferometric Monitoring of Gamma-Ray Bright Active Galactic Nuclei II: Frequency Phase Transfer // JKAS  -- 2015. -- 48, 237A
\bibitem{6}
{\it Hodgson J.A. et al.} The Automatic Calibration of Korean VLBI Network Data // JKAS  -- 2016. -- 49, 137H
\bibitem{7}
{\it Hodgson J.A. et al.} Location of Gamma-ray emission and magnetic field strengths in OJ 287 // arXiv  -- 2016. -- 1607.00725
\bibitem{8}
{\it Nagai H.. et al.} Nature of radio feature formed by re-started jet activity in 3C 84 and its relation with {$\gamma$}-ray emissions // Astronomische Nachrichten  -- 2016. -- 337, 69



\end{thebibliography}
\end{document}